# A Statistical Index for Early Diagnosis of Ventricular Arrhythmia from the Trend Analysis of ECG Phase-portraits

Grazia Cappiello[a], Saptarshi Das[a]*, Evangelos B. Mazomenos[a], Koushik Maharatna[a], George Koulaouzidis[b], John Morgan[b,c], and Paolo Emilio Puddu[d]

*a) School of Electronics and Computer Science, University of Southampton, Southampton SO17 1BJ, UK*

*b) Wessex Cardiothoracic Unit, University Hospital Southampton NHS Foundation Trust, Tremona Road, Southampton SO16 6YD, UK*

*c) School of Medicine, University of Southampton, Southampton SO17 1BJ, UK*

*d) Department of Cardiovascular Sciences, Sapienza University of Rome, Viale del Policlinico 155, I-00161 Rome, Italy*

**Authors' Emails:**

graziacappiello@libero.it (G. Cappiello)

sd2a11@ecs.soton.ac.uk, s.das@soton.ac.uk (S. Das*)

ebm@ecs.soton.ac.uk (E.B. Mazomenos)

km3@ecs.soton.ac.uk (K. Maharatna)

geokoul@hotmail.com (G. Koulaouizidis)

jmm@hrclinic.org (J. Morgan)

paoloemilio.puddu@uniroma1.it (P.E. Puddu)

**Corresponding author's phone number:** +44(0)7448572598

**Abstract:**

In this paper, we propose a novel statistical index for the early diagnosis of ventricular arrhythmia (VA) using the time delay phase-space reconstruction (PSR) technique, from the electrocardiogram (ECG) signal. Patients with two classes of fatal VA - with preceding ventricular premature beats (VPBs) and with no VPBs have been analysed using extensive simulations. Three subclasses of VA with VPBs *viz.* ventricular tachycardia (VT), ventricular fibrillation (VF) and VT followed by VF are analyzed using the proposed technique. Measures of descriptive statistics like mean ($\mu$), standard deviation ($\sigma$), coefficient of variation ($CV = \sigma/\mu$), skewness ($\gamma$) and kurtosis ($\beta$) in phase-space diagrams are studied for a sliding window of 10 beats of ECG signal using the box-counting technique. Subsequently, a hybrid prediction index which is composed of a weighted sum of $CV$ and kurtosis has been proposed for predicting the impending arrhythmia before its actual occurrence. The early diagnosis involves crossing the upper bound of a hybrid index which is capable of predicting an impending arrhythmia 356 ECG beats, on average (with 192 beats standard deviation) before its onset when tested with 32 VA patients (both with and without VPBs). The early diagnosis result is also verified using a leave out cross-validation (LOOCV) scheme with 96.88% sensitivity, 100% specificity and 98.44% accuracy.

## 1. Introduction

Early diagnosis or prediction of VA may allow clinicians sufficient time to intervene for stopping its escalation causing Sudden Cardiac Death (SCD) and thus is an active research area in the field of cardiology. Over the decades the main emphasis has been put on studying the Heart Rate Variability (HRV) as a possible marker for the early diagnosis of VA [1]. Recently it was found that HRV increases two hours before the onset of arrhythmia [2]. In the same work the authors found the beat-to-beat oscillations of T-wave amplitudes increase before the onset of VA. Despite these findings it is difficult to derive a temporal relationship of these markers unequivocally to the onset of VA. In this paper, we have attempted to approach the problem of short-term predictability of VA from a dynamical system theory perspective and through statistical analysis of continuous history of ECG data from 34 patients. The analysis investigates whether it is possible to derive a statistical index, crossing of which above a certain threshold can be considered as an early diagnosis of impending VA before its onset. Such an index may help for stratifying short-term risk of arrhythmia along with the other clinical markers mentioned earlier.

From the perspective of dynamical systems theory, the heart in healthy condition may be considered as a system maintaining constant phase relationship between electrical activities occurring at its different parts leading to an overall synchronised operation. During arrhythmia this coherent phase relationship is disrupted, resulting in a chaotic rhythm. However, such a chaotic rhythm is very unlikely to be set suddenly without approaching the bifurcation point (the onset of arrhythmia here) over a finite time window. Therefore, it may be reasonable to assume that the manifestation of VA in essence is a cumulative effect of an incremental temporal drift of phase relationship between electrical activities occurring at different parts of the heart leading to a desynchronisation phenomenon over a finite time frame. Another important point to note is that the ECG of a patient with tendency of arrhythmia often shows the presence of single or multiple VPBs which can possibly be considered as a compensatory mechanism that the human body takes for mitigating this gradual desynchronisation process [3]. Once such mechanism fails, the arrhythmia occurs which is manifested as random fluctuations in the ECG traces (mostly polymorphic VT) instead of its typically well-behaved P-QRS-T pattern.

Although in principle such a temporal phase drift is expected to be imprinted in the ECG time-series as changes in inter-beat interval, its value could be small enough to be detected by visual inspection of ECG alone until at the very last phase. The cumulative effect of increasing inter-beat desynchronisation pushes the heart more towards the verge of chaos or arrhythmia when the ECG beat patterns completely changes to VT/VF. Had such desynchronisation process been detected early, it could serve as a *predictor* of possible impending VA. Phase-space reconstruction (PSR) or time delay embedding [4], [5] is a technique widely used in the field of nonlinear dynamics for detecting such small desynchronisation phenomena in a time-series data which is often indistinguishable by simple observation. Therefore such technique when applied on ECG time-series may have potential for detecting the gradual phase desynchronisation leading to the arrhythmia. Although several studies like [6–8] have established that the PSR has better capability to transform the ECG time series in the form of images where such desynchronisation can be better understood, still in standard practice in clinical cardiology is mostly based on the observation of ECG beats along with other time domain parameters and not the phase portraits which could be a popular way of diagnosis pathological ECGs in future.

In PSR technique a dynamic system's trajectory is reconstructed by plotting the original signal and its delayed versions along mutually orthogonal axes of the co-ordinate system. This gives a closed contour for a periodic signal representing a limit cycle for the regular oscillations of the system under investigation. This technique is widely used in detection of chaos or deviation from constant frequency oscillations, if the system is considered to be noise-free. In healthy heart condition, within a small time window, the consecutive ECG beats can be considered as an almost periodic waveform and therefore the phase-space analysis of it would produce an almost closed contour [9]. In the presence of certain desynchronisation process, these closed contours or trajectories will start to spread depending upon the amount of desynchronisation in the system and therefore counting the number of trajectories and their statistical variations over time may give a quantitative metric for temporal dynamics of the underlying desynchronisation process. In this paper, we exploit this philosophy for investigating formulation of a possible short-term prediction index of VA.

The phase-space analysis technique has been used successfully over the years for detecting VT and VF [6–8], as biometric for human identification [10], coronary occlusion [11], heartbeat classification [12], detection of ECG fiducial point [13], fetal ECG monitoring [14], [15], spatial analysis using vectorcardiogram [16], analyses of QRS-complex time series [17], heart rate detection [18], and for understanding heart rhythm dynamics [19], [20]. Apart from the above mentioned works on application of ECG phase portraits for analysing abnormal heart conditions, there were attempts for the prediction of irregular heart conditions e.g. study of VF using multi-parameter analysis [21], AF prediction from heart rate variability and ECG signals [22], [23], risk of arrhythmia in post Myocardial infarction [24], [25], prediction of defibrillation success [26], prediction of VF duration using angular velocity [27] etc. among many others. In addition, among other popular nonlinear dynamical measures, the period-doubling bifurcation [28], correlation dimension [29] and complexity measure [30] have also been applied to detect fatal arrhythmia like VF. However the major emphasis in all of these works is the *detection* (not *early diagnosis* or *prediction*) of abnormal heart conditions after the anomalous behaviour in the heart or arrhythmia has been manifested. Although there are ECG PSR based methods to detect arrhythmic events once it is manifested, but there is almost no literature in similar works on early diagnosis of impending arrhythmia.

The aim of this work is to formulate a regularised statistical index that may *predict or* give an *early diagnosis* of the impending VA, specifically VT and VF, before its onset by crossing a certain threshold. The proposed index has been formulated from the statistical trend analysis of the trajectories of PSR for long ECG time-series history to capture the underlying temporal desynchronisation process. The PSR when applied on ECG data up to the onset of arrhythmia, results in 2-D images, describing the system trajectories which are analysed using the well-known box-counting technique [4], [5]. To achieve our goal, we first study four descriptive statistics as the moments of order $1 - 4$ *viz.* the mean ($\mu$), variance ($\sigma$), skewness ($\gamma$) and kurtosis ($\beta$), corresponding to the temporal evolution of the trajectories in healthy and VA populations – both cohorts are of size 32. Two main classes of arrhythmic patients were considered – patients having VA without VPBs and patients having VA with one or multiple VPBs. In the latter class; three subclasses were considered, patients with – VF, VT and VT followed by VF. In addition to the above mentioned statistical parameters, the coefficient of variation ($CV = \sigma/\mu$) has been introduced as another potential statistical measure which represents the spread of the trajectories normalised by its mean. The statistical variation of the number of trajectories resulted from phase-space reconstruction is computed using the box-counting technique [4], [5]. The present paper thoroughly analyses the first result of such early diagnosis of VA from the statistical trends of ECG phase portraits [31].

The rest of the paper is organized as follows: section 2 describes the basics of phase-space reconstruction, box-counting and adopted signal and image processing techniques and in section 3 the strategy for analysing the healthy and arrhythmic ECG signals is elaborated. Section 4 describes the statistical analysis of the phase-space diagrams and the proposed index is formulated in section 5. The conclusions are drawn in section 6 with a discussion on future scopes of research.

## 2. Theoretical formulation

### 2.1. Basics of phase-space reconstruction and box-counting

Study of phase space reconstruction is commonly done in the field of nonlinear dynamical (especially chaotic) systems where the system is considered to be deterministic and generally represented by a set of ordinary differential equations. The Takens' theorem proves that a sufficiently large time delayed versions of only one measured state $x_n$ can be used for the reconstruction of the underlying dynamics of the system in the state-space or phase-space [5]. In other words, from a measured discrete time series data $x_n$, if the delay $d_e$ is sufficiently large, the evolution of $\{x_n, x_{n-1}, x_{n-2}, \cdots, x_{n-d_e}\}$ will be the same as the dynamics of the underlying higher dimensional physical system. The time delay $d_e$ is chosen in such a way that the phase space trajectories have the maximum span [5].

Once the phase space trajectories are constructed, it is necessary to analyse their statistical behaviour, e.g. the number of trajectories and their spread etc. The well-known technique of box-counting [4], [5] can be applied for that purpose. It is in general used for studying the fractal dimension of a graph. In the present methodology, the entire phase-space diagram is represented as a 2-D image (typically called phase portraits) of $N \times N$ pixel, where $N$ is an integer. The pixels through which at least one trajectory has passed are considered as black boxes ($n_b$) and the others are considered as white boxes ($n_w$). The degree of complexity or chaotic dimension of the phase portrait is represented using a metric, defined as the ratio of the number of black boxes ($n_b$) and the total number of pixels ($n_w + n_b = N^2$) in the portrait. The concept originated from the box-dimension or Hausdorff index of fractal images which is the ratio of the logarithm of number of black pixels and negative of logarithm of pixel length as the size of the boxes approaches zero [5]. But here, we did not decrease the size of boxes and carry out box-counting since the study is not intended to investigate fractal behaviour of the problem but to quantify the spread of the phase space images of the windowed ECG. The ECG phase-portrait box-counting has been shown as an efficient *detection* tool for VF and VT in [6], [7], which has been extended in this paper as a possible *prediction* or *early diagnosis* tool by looking at the different statistical trends of the number of black boxes visited. Since arrhythmia is somewhat similar to the appearance of chaos in a dynamical system, hence it should get reflected in the phase space box-counting even if with a fixed size of the boxes as a quantification of the spread of the ECG phase portraits. Therefore the concept of PSR of ECG and its box-counting has been used only to quantify when the desynchronisation exceeds the '*safe level*' which can further lead to impending arrhythmia and not to investigate fractal behaviour. Although the reconstructed ECG phase portrait is a low dimensional projection of the original high-dimensional attractor but we here explore how it serves the purpose of early diagnosis of impending arrhythmia.

## 2.2. ECG selection and adopted signal processing techniques

Publicly available ECG databases from Physionet [32] have been used in the present study. For the analysis of healthy and arrhythmic ECGs, 32 subjects were selected both from the PTB diagnostic database (PTBDB) and Creighton University Tachyarrhythmia database (CUDB). The 32 ECG traces from CUDB (out of total 35 entries in the database) could be analysed due to their unambiguous interpretation (i.e. less corrupted with artefacts) and the severity of arrhythmia. In addition, the arrhythmic signals are categorized in two major classes - one without VPBs (5 patients) and the other with single or multiple occurrences of VPBs (27 patients). Within the class of VA with VPBs, we have further considered three subclasses *viz.* four patients with VF, 13 patients with VT, and 10 patients with VT which translates to VF over time. Here, most of the patients have sufficiently long healthy-looking ECG heartbeats (with clearly distinguishable P-QRS-T waves) with occurrences of VPBs before the onset of arrhythmia, except the VA with no VPBs class where the signals are relatively short in length for some patients. All the selected ECG traces are clinically annotated to identify the healthy-looking beats, VPBs and the onset of arrhythmia.

For the PSR, we selected a window of 10 successive ECG beats. Since the PSR yields best results if the signals are noise-free, all the ECG signals were filtered using a fourth-order Butterworth high-pass digital filter with a cut-off frequency of 1 Hz, to eliminate the drift or baseline wandering, followed by a low-pass filter with a cut-off frequency of 30 Hz to eliminate frequencies higher than that, which mainly consist of measurement noise [33]. Each ECG window of 10 beats was normalized using (1) in order to ensure that all values are within (0, 1) as in [10]:

$$\tilde{x}(t) = \left(x(t) - x_{\min}\right) / \left(x_{\max} - x_{\min}\right) \tag{1}$$

In (1), $x_{\min}$ and $x_{\max}$ are the minimum and the maximum values of the filtered ECG data $x(t)$, while $\tilde{x}(t)$ is its normalized form. In Physionet, while the ECG signals from the PTBDB are sampled at 1 KHz, the CUDB signals are originally sampled at 250 Hz [32]. To bring them to a uniform platform we interpolated the CUDB signals to 1 KHz. We took the approach of oversampling or interpolating the CUDB signals, instead of down-sampling the PTBDB signals, since it is well-known that interpolation increases the quality of the signal and reduces the effect of noise, thus improving the signal to noise ratio (SNR). Many ambulatory ECGs which often have low sampling rate is interpolated to improve the online calculation of heart rate and other clinical features, as reported in [34], [35]. Also, for a few patients from CUDB database some artefact corrupted ECG traces were neglected. Following the exploration reported in [7], [10] for the optimum PSR of ECGs, we added 20 samples or 20 ms of delay for the phase-space reconstruction of the filtered and normalized ECG signals, since amongst various other embedding delays this particular choice gives good person-centric characterization. Although there could be other methods to select the optimum time delay embedding e.g. autocorrelation, mutual information, approximate period, generalized embedding lags etc. [4], [5], we base our analysis on the choice reported in [10].

A total of 10 trajectories are obtained from a window of 10 ECG beats in a 2-D phase-space diagram which is resized and then exported as a high-resolution gray-scale image of pixel size 1024×1024. The box-counting method for analysing the statistical properties of this phase portrait was carried out using Matlab where the black and white pixels were assigned a "0" and "1" value respectively. Subsequently the number of black pixels was counted as they indicate the measure of spread of the trajectories and hence the underlying desynchronisation phenomenon. For statistical analyses of the trajectories 25 such phase portraits were used in a sliding window fashion for deriving different run-time measures $\{\mu, \sigma, \gamma, \beta, CV\}$ of the number of black boxes visited. The number of phase portrait windows was set to a higher value (25), since in order to capture the different statistical moments, the histogram of the number of black boxes visited needs to be constructed in sufficient detail which needs larger number of data–points. These five statistical measures of the number of black boxes (i.e. $\mu$, $\sigma$, $\gamma$, $\beta$, $CV$) visited and their combinations are expected to characterise the underlying desynchronisation phenomenon in sufficient detail yielding an early VA diagnosis tool.

## 3. Methodology for statistical analysis of ECGs

### 3.1. Methodology adopted for the healthy subjects

For the analysis of healthy ECG signals from the PTBDB the boundaries of each heartbeat have to be detected reliably. For this purpose, we used the automated time domain morphology and gradient (TDMG) algorithm [36], based on a combination of extrema detection and slope information, using adaptive thresholding as shown in Fig. 1(a). The vertical lines in Fig. 1(a) shows the start and end boundaries of the automated ECG segmentation using TDMG algorithm.

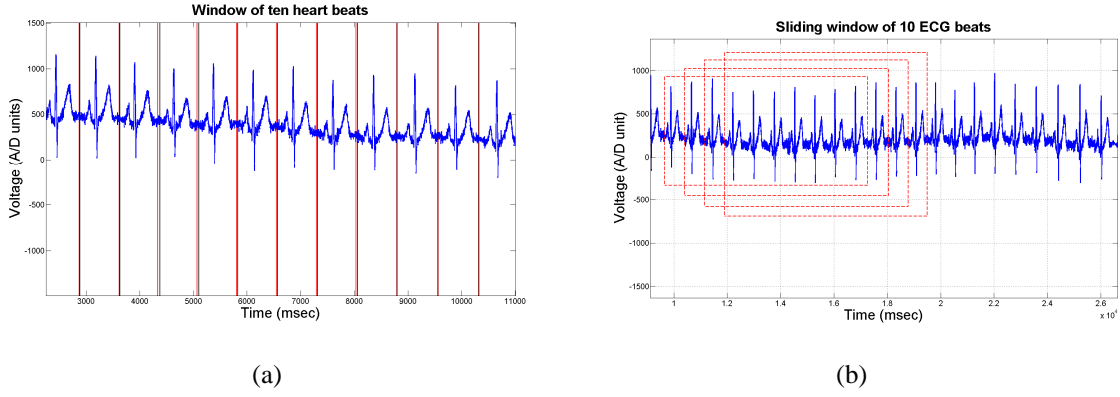

(a)                                          (b)

Fig. 1.(a) The extraction of the beginning and the end instants of each heart beat from an ECG signal using TDMG algorithm, (b) Sliding window of 10 consecutive heart beats with 9 beats overlap.

Once the beginning and end of each ECG beat are identified, a vector $y$ was constructed as in (2).

$$y = \left\{ t(i) \right\}, i \in \left[ 1, 2, \cdots, (n+1) \right] \qquad (2)$$

where, $t(i)$ is the beginning instance of $i^{th}$ or end instance of $(i-1)^{th}$ heartbeat. This is done for all the beats until $t(n+1)$ which is the end instance of the $n^{th}$ heartbeat. Using a sliding window of 10 consecutive ECG beats with 9 beats overlap at a time as shown in Fig. 1(b), a sliding-window matrix ($M$) was constructed from the vector $y$ described in (3).

$$M = \begin{bmatrix} t(1) & t(2) & \cdots & t(11) \\ t(2) & t(3) & \cdots & t(12) \\ \vdots & \vdots & \ddots & \vdots \\ t(n-9) & t(n-8) & \cdots & t(n+1) \end{bmatrix} \qquad (3)$$

An example of filtered and normalized delayed versions of the ECG signal (as described in section 2) have been shown in Fig. 2. Plotting the locus of the discrete samples for 10 consecutive ECG beats, by considering the normalized signal $\tilde{x}(t)$ and the delayed signal $\tilde{x}(t-20)$ in a unit of millisecond along the two orthogonal axes respectively, yielded the phase portrait shown in Fig. 2 containing 10 trajectories. It is observed that all the trajectories in the phase-space are close to each other and lie almost within an annular band. This indicates towards the regularity of oscillation in healthy ECG and also indicates that the number of boxes visited for healthy ECG phase portraits can be expected to lie within a small range.

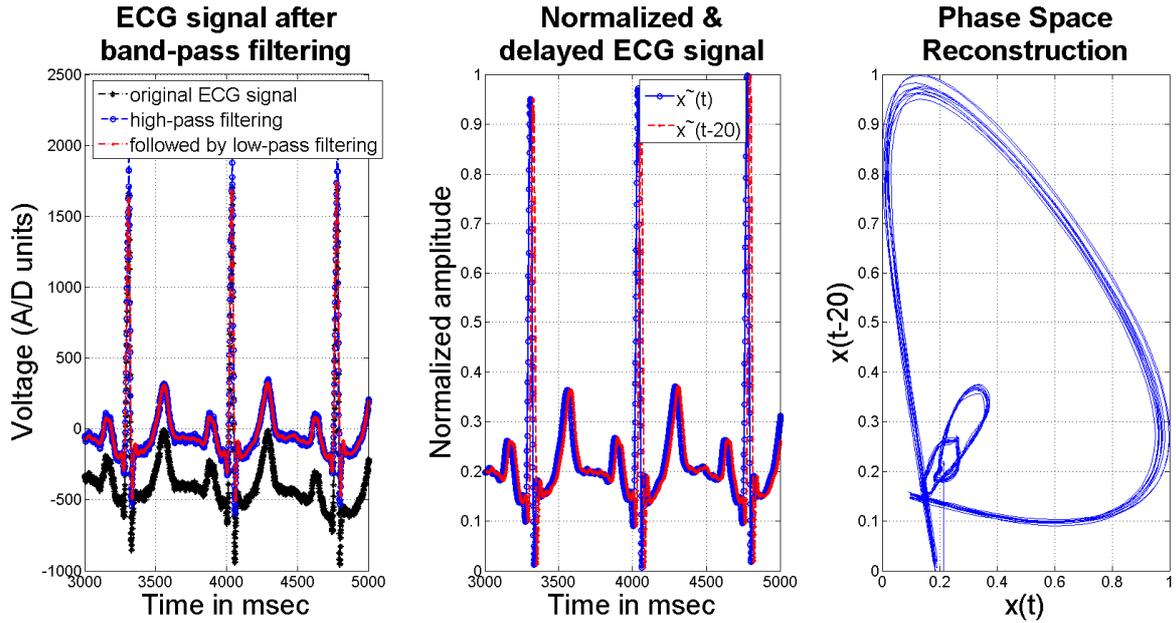

Fig. 2. Phase-space reconstruction of ECG beats from filtered and normalized delayed signal.

### 3.2. Methodology adopted for the patients affected by ventricular arrhythmia

Similar analyses have been carried out for the patients with VA as well. Due to the variability of the heart beat duration and beat morphology over time in arrhythmic subjects with frequent occurrences of stretched and squeezed beat-length, the beginning and end instances of each heart beat were annotated and counted manually by the clinicians, including the occurrence of VPBs, due to the unavailability of any reliable automated beat segmentation algorithm for arrhythmic ECG, unlike the same for healthy ECG [36]. The annotated beats are then transformed in the phase portraits which may include VPBs or beats with a different morphology or varying duration with elongated/squeezed beat length. Also, the mean beat duration was calculated for the normal ECG beats before the occurrence of the first VPB and has been used to locate the boundaries even in the case of double, triple and other family of VPBs during clinical annotation phase. This creates a sliding window of 10 consecutive heart beats, including VPBs when they occur besides healthy-looking beats. In order to investigate the periodic nature of the heart rhythm for VA subjects, we considered a window of 10 ECG beats instead of taking fixed time interval for the analysis. Because a fixed time window may lead to consideration of half a beat or some extra beat which may break the desired periodicity leading to spurious result. Therefore, our analysis considers 10 beat as the window for PSR and is consistently applied on both the groups – healthy and arrhythmic subjects.

In addition, some artefact corrupted parts of the ECG were neglected and the analysable parts have been extracted by concatenating the clean part of the signals before and after the artefact, otherwise it may drastically increase the number of visited boxes. Even in standard practice in cardiology, the diagnosis of arrhythmia is done in time domain, after removing all artefacts from the raw ECG data. Before and after the artefact corrupted ECG beats, the pre-processed ECG looks like normal periodic pacing or sinus rhythm. Several studies like [37], [38] has suggested to neglect the artefact corrupted parts of the ECG time series and use the processed data for investigation of arrhythmia. In fact, the presence of artefacts in healthy ECG signal cannot be removed using any standard filtering since the frequency spectrum of the ECG and artefact is mostly overlapped [39] and while removing artefacts by filtering, it may remove essential information of the ECG as well. For this reason, removal of the corrupted portion of ECGs has been verified manually by the clinicians, involved in the study, where the ECG beats are mostly masked by excessive artefacts. Moreover, the number of boxes does not increase by removing artefacts and concatenating the beats segments since the artefact free data contains healthy looking ECG beats. The artefact corrupted beats are chopped off in such a way that it has minimal discontinuities in the ECG time series.

Now, using the same method of phase-space reconstruction, from pure visual inspection, even for arrhythmic cases before the onset of arrhythmia (Fig. 3(a)), it is apparent that some parts of the ECG beats have resemblance with healthy ECGs as in Fig. 2(b), whereas the other parts (particularly involving presence of VPB) are markedly different (Fig. 3(b)). This motivates us for studying various statistical parameters of the entire ECG time traces in terms of the number of black boxes visited in the ECG phase portraits to distinguish between healthy and 'going to be abnormal' (although morphologically healthy-looking) ECG beats before the onset of arrhythmia. Once the arrhythmia occurs, e.g. in the case of VT and VF as depicted in Fig. 3(c) and Fig. 3(d) respectively, the phase portraits constructed using an equivalent length of the ECG heart beats as done for initial healthy parts before arrhythmia, show large chaotic motions in all trajectories indicating towards a higher number of black boxes visited compared to the previous cases.

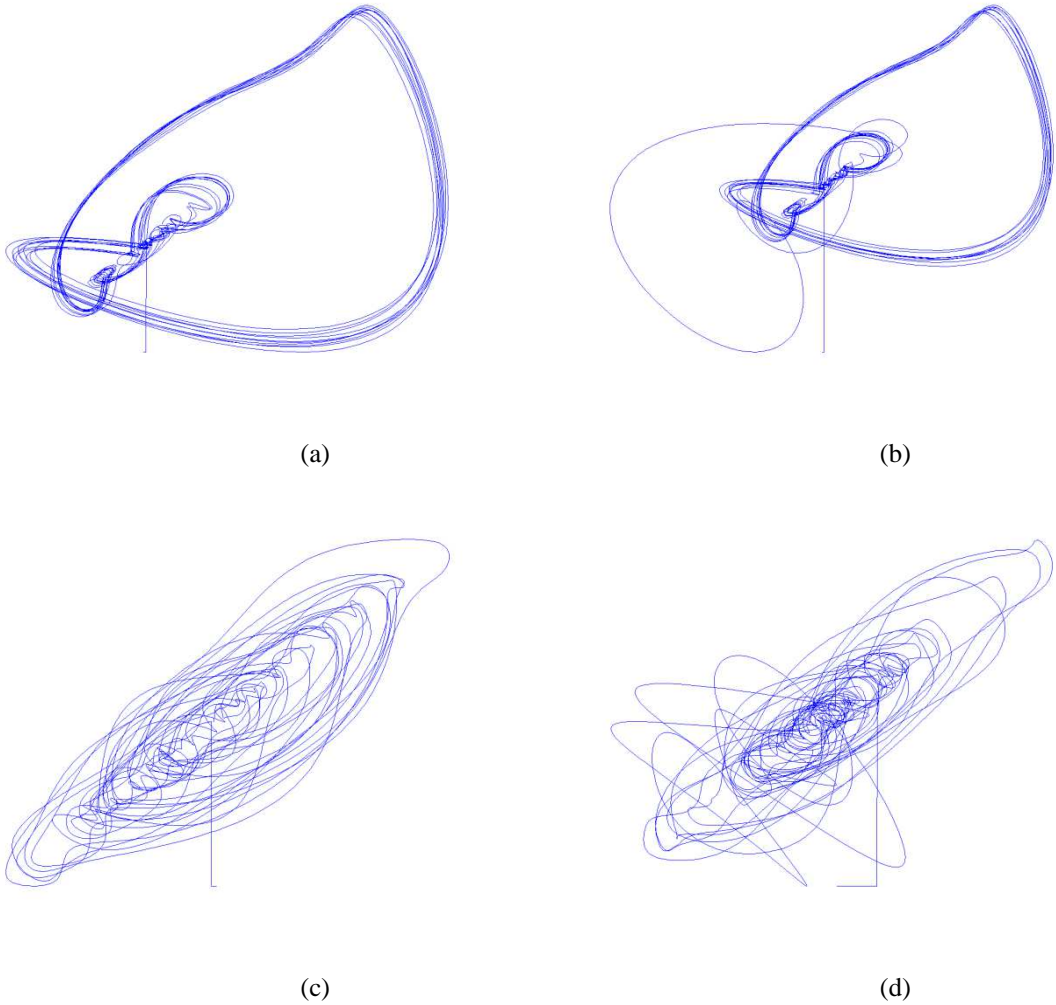

(a)                                             (b)

(c)                                             (d)

Fig. 3. Black and white image of a window of ten beats for (a) normal looking ECG before arrhythmia, (b) with VPB before arrhythmia, (c) VT, (d) VF.

## 4. Statistical analyses of ECG phase portraits

The number of the black boxes visited ($n_b$) by 10 consecutive ECG beats in the phase-space diagram, obtained by the method described in section 3 in essence gives a measure of regularity of the oscillation in ECG waves. To characterise the degree of regularity of such oscillations and hence for identifying any underlying desynchronisation phenomenon, four first-order statistics related moments of order 1-4 i.e. $\{\mu, \sigma, \gamma, \beta\}$ of the number of black boxes visited by the trajectories in each image were studied. Here, the $\{\mu, \sigma\}$ are given by the first and second central moment of the number of black boxes, whereas the $\{\gamma, \beta\}$ represent the third and fourth standardized moments of $n_b$ as in (4).

$$\mu = E[n_b], \sigma = \sqrt{E[n_b - \mu]^2}, \gamma = E\left[\frac{n_b - \mu}{\sigma}\right]^3, \beta = E\left[\frac{n_b - \mu}{\sigma}\right]^4 \qquad (4)$$

The statistical moments have been computed in a sliding window fashion (i.e. the run-time statistics) considering 25 phase-portrait images as the window which is moved stepwise with an overlap of 24 images, thus taking a new image at each step and leaving the oldest one. The choice of the window for the phase portraits was made sufficiently long (in our case 25) so that it allows to captures all intricate details of the box-counting histograms in order to account for the higher order statistical moments. In general for a window containing $W_1$ number of ECG beats, the $n^{th}$ phase portrait represents the phase-space behaviour from $n^{th}$ to $(n + W_1 - 1)^{th}$ number of consecutive ECG beats. Therefore if $W_2$ such phase portraits is used for obtaining the statistics for $\{\mu, \sigma, \gamma, \beta\}$, then it results into the $m^{th}$ data point (typically termed as $m^{th}$ image) that captures the statistical trends for the above parameters between the $m^{th}$ and $(m + W_2 - 1)^{th}$ number of phase portraits. Considering $W_1 = 10$ and $W_2 = 25$ in the present case, the $m^{th}$ data point in the trends of the above mentioned statistical measures would contain the information of $m^{th}$ to $(m + W_1 - 1 + W_2 - 1)^{th} = (m + 33)^{th}$ consecutive ECG beats. In our analysis we have plotted the run-time statistics of the four parameters stated above with respect to the phase portrait window number ($m$) to show the temporal variations of these parameters. Considering the nominal heart rate (HR) in beats/min the relationship between $m$ and the absolute time $t$ in sec can be computed as (5).

$$t = 60(m + W_1 + W_2 - 2)/HR \qquad (5)$$

The calculation of the mean HR of the arrhythmic subject was done by averaging the number of ECG beats per minute over the whole length of the signal for that particular arrhythmic subject. We used the manual annotations of the clinician including different abnormalities like VPBs, different morphology or varying duration with elongated/squeezed beat length. The calculation of the mean beat duration was particularly necessary to detect the boundaries of the ECG beats in the cases of single, double, triple VPBs or short episode of VTs along with healthy beats. This way it was possible to generate the phase portraits with an approximate length of 10 healthy-looking ECG beats for that particular arrhythmic subject.

In Fig. 4, the phase portrait box-counting mean ($\mu$) and standard deviation ($\sigma$) trends for one representative example from a healthy (PTBDB) and an arrhythmic (CUDB) patient are shown. It is evident that for the healthy subject, the $\mu$ and $\sigma$ trends are almost uniform throughout the time trace. On contrary, the arrhythmic subject shows sudden increase in both $\mu$ and $\sigma$ at the last stage of the trend values indicating the onset of arrhythmia. It is evident that the already started gradual desynchronisation, ultimately manifested as the onset of arrhythmia [40], can be detected from $\mu$ and $\sigma$ trends by observing the sudden jump in the box-counting. However before the arrhythmic region, the trends of $\mu$ and $\sigma$ do not differ considerably from those of the healthy subjects. Therefore it appears that although these two parameters could be used for *detecting* arrhythmia once it is manifested, their *predictive* value for early diagnosis of impending arrhythmia – the main target of this work – is rather low.

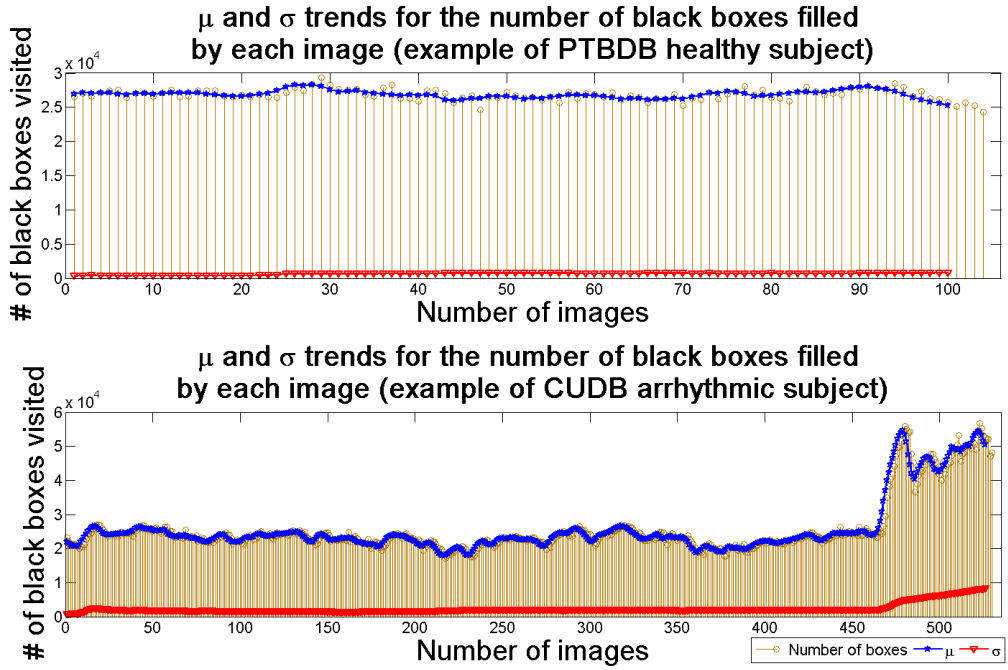

Fig. 4. Box counting $\{\mu, \sigma\}$ trends for healthy and arrhythmic patient respectively.

## 4.1.    Analysis of healthy subjects in PTBDB

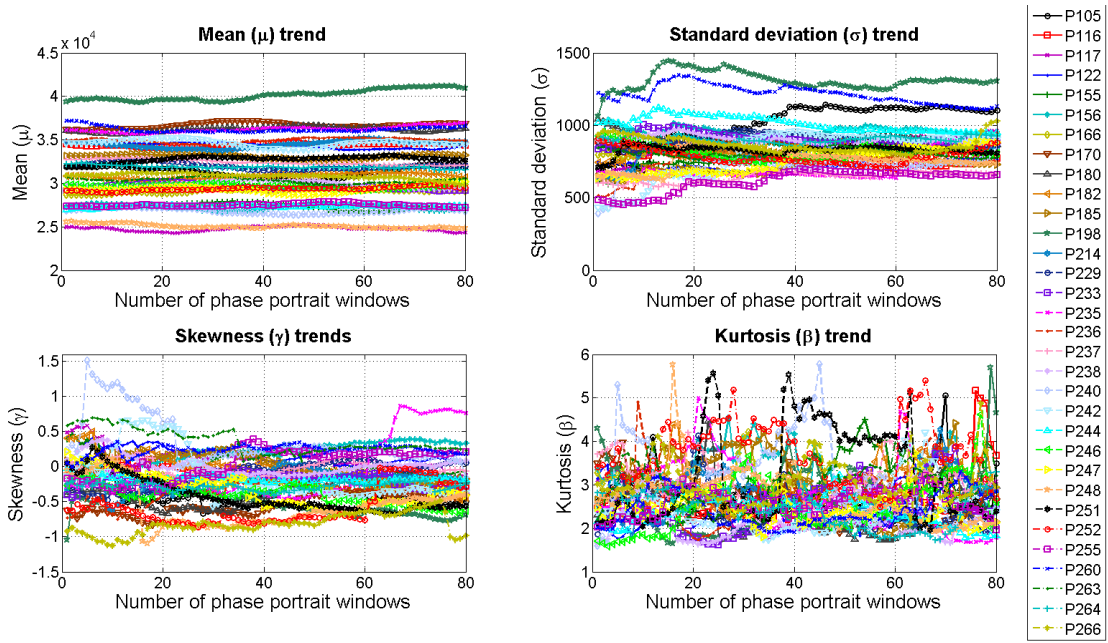

Fig. 5. Trends of four statistical measures $\{\mu, \sigma, \gamma, \beta\}$ of box-counting for healthy subjects.

The trends of all the four moments $\{\mu, \sigma, \gamma, \beta\}$ for 32 healthy subjects are plotted in Fig. 5 with the respective subject numbers of PTBDB indicated in the legend. It is clear that although $\mu$ trend varies slowly, $\sigma$ and $\gamma$ trends vary relatively faster and high fluctuations are observed in the $\beta$ trends. However, as observed before, $\mu$ and $\sigma$ trends may not have high predictive value as individual parameters. This may not be completely surprising given the possibility of significant inter-person variability which is also evident from Fig. 5. Therefore a more pragmatic approach would be to examine the relative spread of the trajectories which can be captured by introducing the coefficient of variation $CV = \sigma/\mu$, which in essence computes the trajectory spread normalised by the mean. The $CV$ trends for health subjects are shown in Fig. 6. Quite interestingly from Fig. 6 it is observed that *CV* has less inter-person variability and is always bounded by an upper limit of $CV = 0.05$. This implies that in the $\mu$–$\sigma$ plane, the bound for time evolution of *CV* trends can be represented by a straight line $\sigma = 0.05\mu$ below which lies the '*safe region*' for the heart beat synchronisation, indicating the healthy condition of heart. A similar behaviour can also be observed from Fig. 5 for *β*. Since *β* in essence characterises how much peaked the distribution of the number of black box visited is and any change in that distribution is reflected as a spike in its trend. Although such spikes are evident in Fig. 5 for all the subjects, the important point to note here is that all the magnitudes of such peaks are bounded by $\beta < 6$. Therefore, from the above simulations, two upper thresholds have been found for the healthy subjects i.e. $CV_{th} = 0.05$ and $\beta_{th} = 6$, crossing of which may provide an early indication towards gradual increase in desynchronisation amongst the ECG beats.

From the available standard 12-lead ECG of PTBDB, we have consistently chosen the lead-I for all the healthy subjects. For the case of arrhythmia, generally all the leads captures its trace but may slightly differ in morphology and amplitude. Since our work is based on the normalized phase portrait images, it does not get affected by slight change in ECG amplitude but rather depends more on the inter-beat synchronisation and change in underlying dynamics which gets captured in all the ECG leads during arrhythmia.

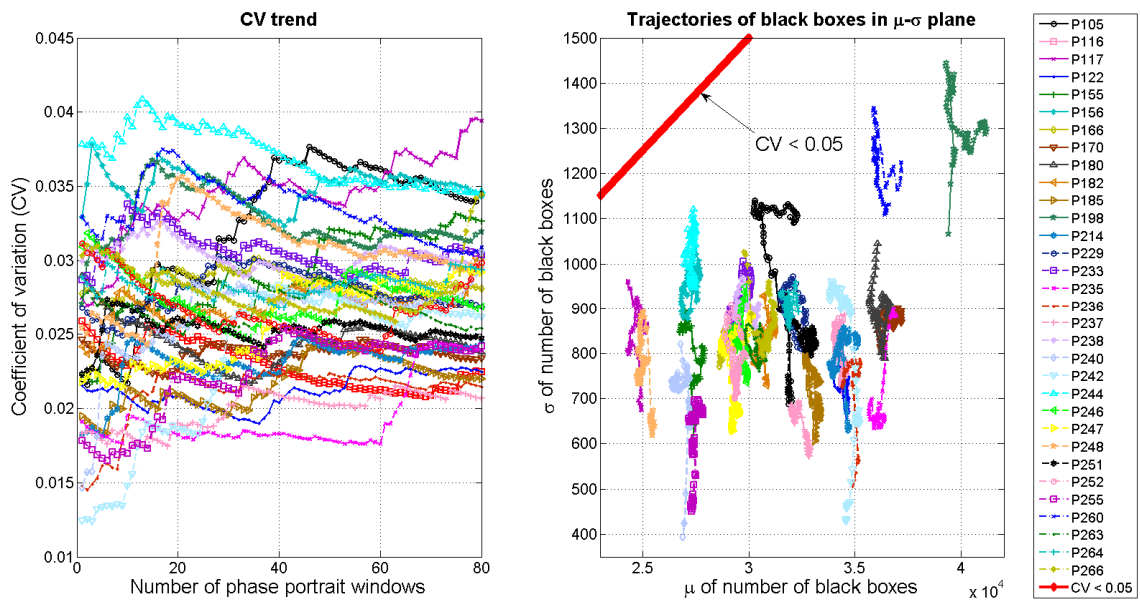

Fig. 6. *CV* trends of box-counting for healthy subjects.

## 4.2. Analysis of arrhythmic subjects in CUDB

As mentioned in earlier sections, the analysable 32 arrhythmic subjects from the CUDB has been categorized in four groups – five patients had VA with no VPBs, four patients were affected by VF, 13 patients by VT, and 10 patients by VT followed by VF. The $\{\mu, \sigma, \gamma, \beta\}$ trends of the number of black boxes visited by each phase-portraits for these different subclasses of VA (i.e. with no VPBs, VT, VF, VT followed by VF) are shown in Fig. 7-10 respectively. In each statistical trend and class of arrhythmia, the respective subject numbers are mentioned in the legend of the Figs. 7-10. The onset of arrhythmia in terms of the number of phase portrait window have been reported in Table 1 for each group of patients.

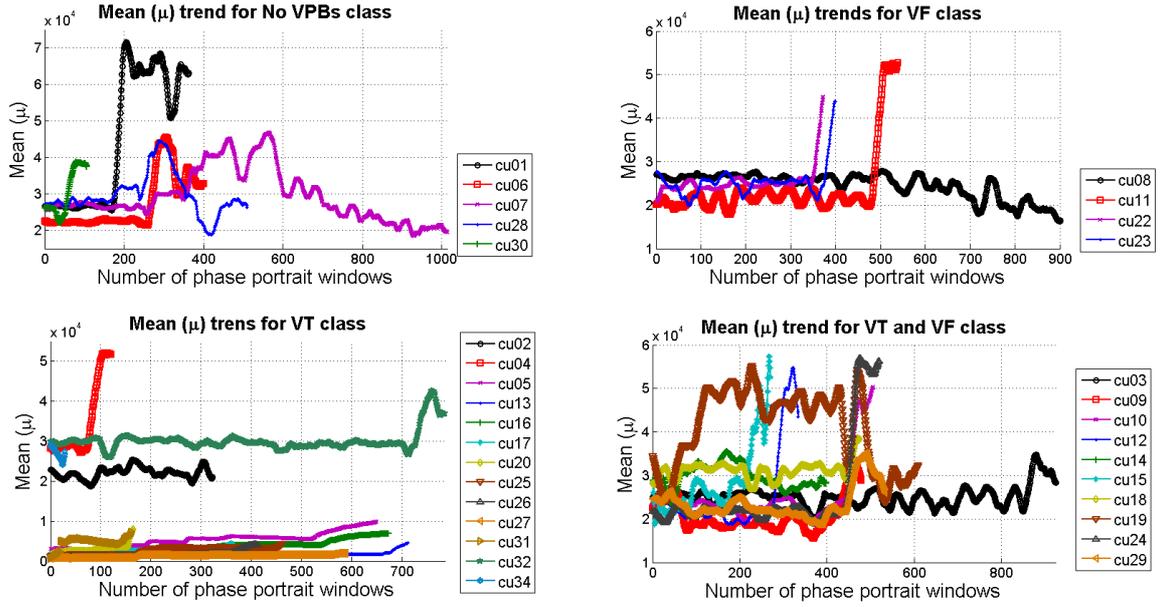

Fig. 7. Mean ($\mu$) trends of box-counting for arrhythmic subjects.

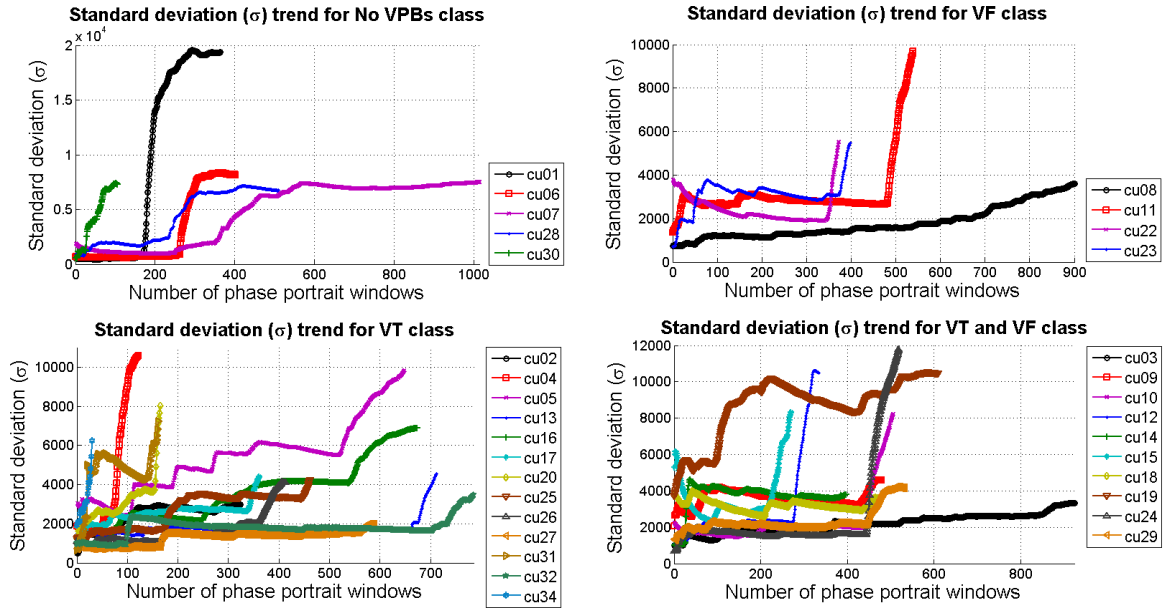

Fig. 8. Standard deviation ($\sigma$) trends of box-counting for arrhythmic subjects.

As expected the number of black boxes visited increases considerably in the $\{\mu, \sigma\}$ trends of the box-counting when arrhythmia occurs and it is related to the irregular behaviour of the trajectories in the phase-space diagrams. In fact sudden fluctuations in the $\{\mu, \sigma\}$ trends are observed when the VPB is encountered, which is different from that of the healthy subjects in Fig. 5. The $\gamma$ trends in Fig. 9 also show sudden increase in its value compared to that for the healthy subjects in the arrhythmic region indicating that it may have high arrhythmia *detection* power. However, before the arrhythmic region its trends are more or less similar to those of the healthy subjects. Therefore like the individual trends of $\mu$ and $\sigma$, its *predictive* power is limited. However a marked difference is observed in the $\beta$ trends for all arrhythmic subjects compared to the healthy subjects. Even before the arrhythmic region, $\beta$ shows spikes that are much higher in magnitudes than those of the healthy subjects. In general, comparing Fig. 5 with Fig. 10 it may be said that for the healthy subjects $\beta$ is bounded by $\beta <$ 6, whereas for arrhythmic subjects $\beta > 6$ in all cases, even before the occurrence of arrhythmia. This shows that $\beta$ may have significant *predictive* value for impending arrhythmia.

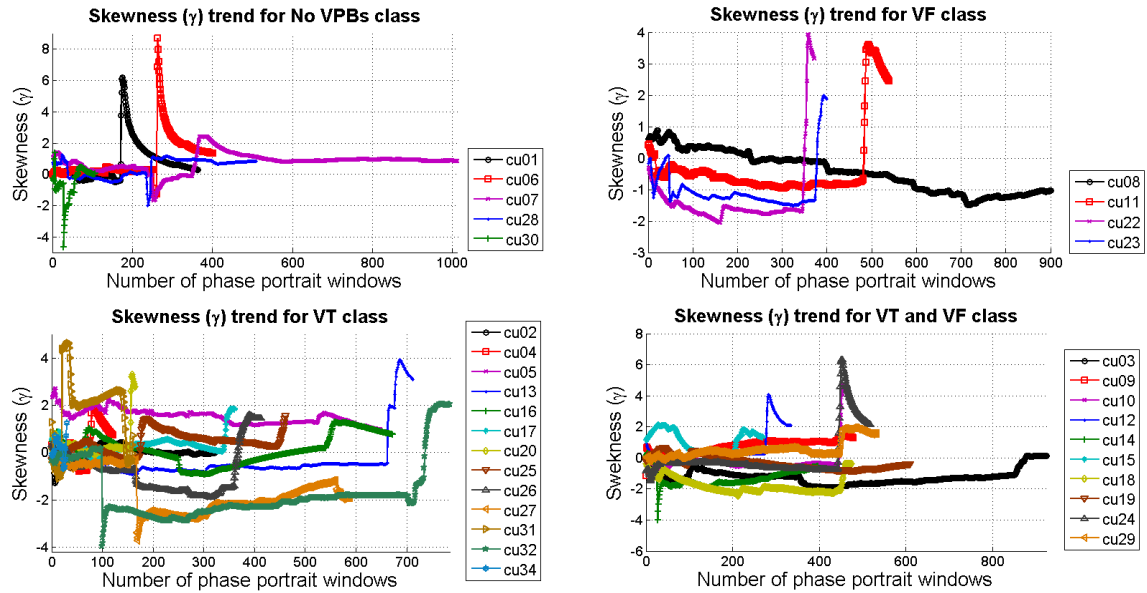

Fig. 9. Skewness ($\gamma$) trends of box-counting for arrhythmic subjects.

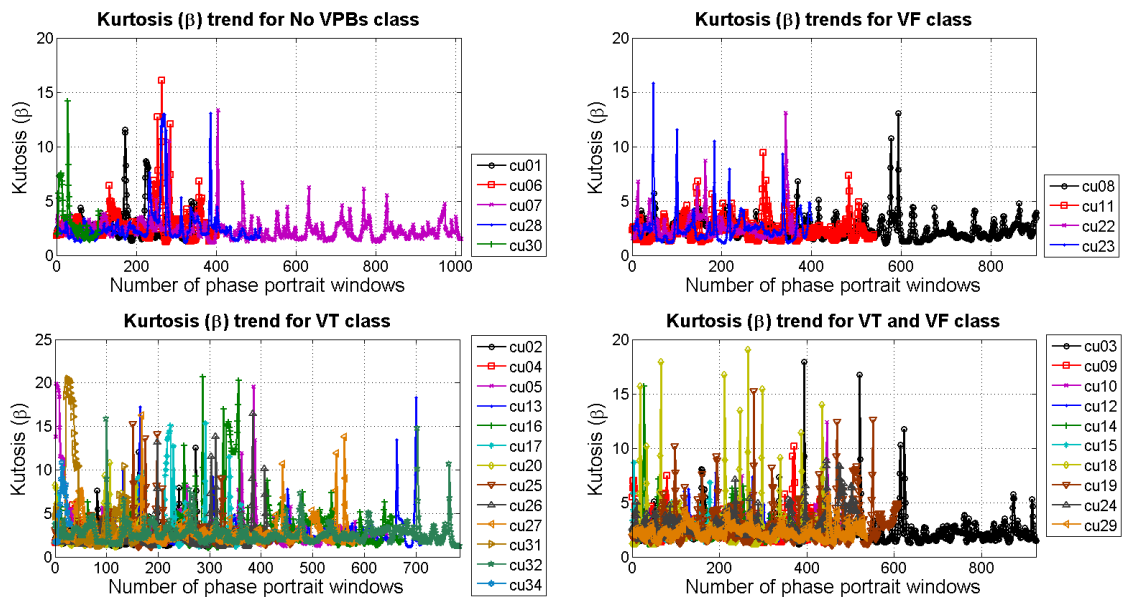

Fig. 10. Kurtosis ($\beta$) trends of box-counting for arrhythmic subjects.

Another interesting observation comes from the analysis of *CV* trends as shown in Fig. 11 and Fig. 12. For the three arrhythmia subclasses *viz.* VF, VT and VT leading to VF – all with VPBs – *CV* trends cross the threshold $CV_{th} = 0.05$, which is the bound of *CV* for healthy subjects, much earlier than the actual arrhythmic event indicating towards its possible predictive property. On the other hand for the cases of VA without VPBs, *CV* trends show similar nature to those in the healthy subjects before the onset of arrhythmia. Therefore it appears that although *CV* trend could be considered as a good predictor for some subclasses of arrhythmia its predictive power for arrhythmia with no VPBs is limited. The predictive capability of the *CV* trends is explored in its trend form in Fig. 11 and also in the $\mu$-$\sigma$ plane (as time evolves) in Fig. 12 where the trajectories of black box crosses the upper threshold of $CV = 0.05$ (in Fig. 6) much earlier than the actual occurrence of arrhythmia.

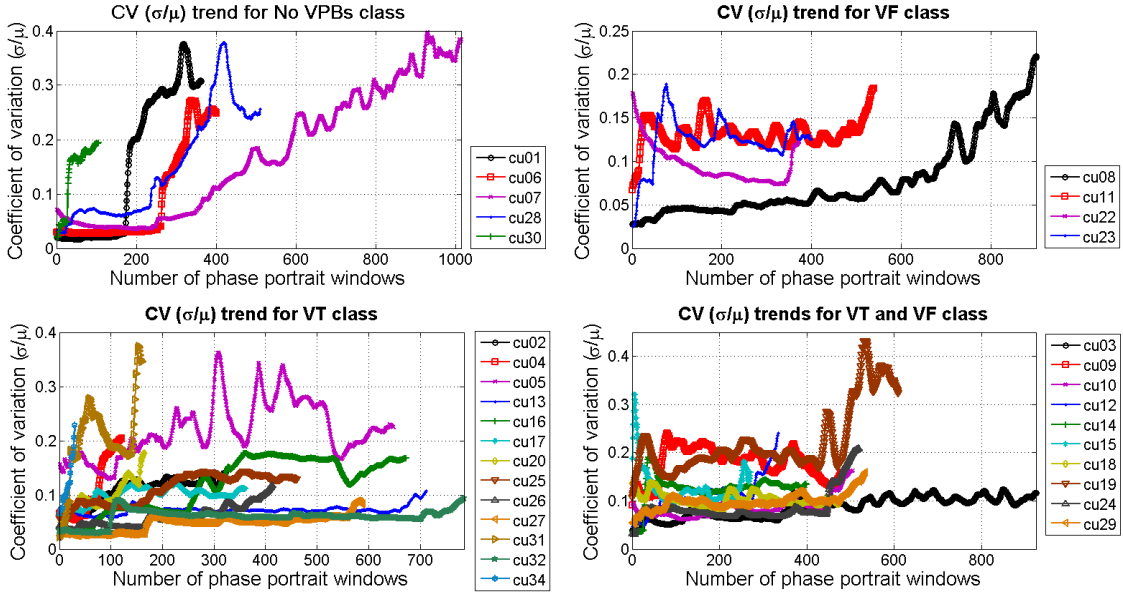

Fig. 11. *CV* trends of box-counting for arrhythmic subjects.

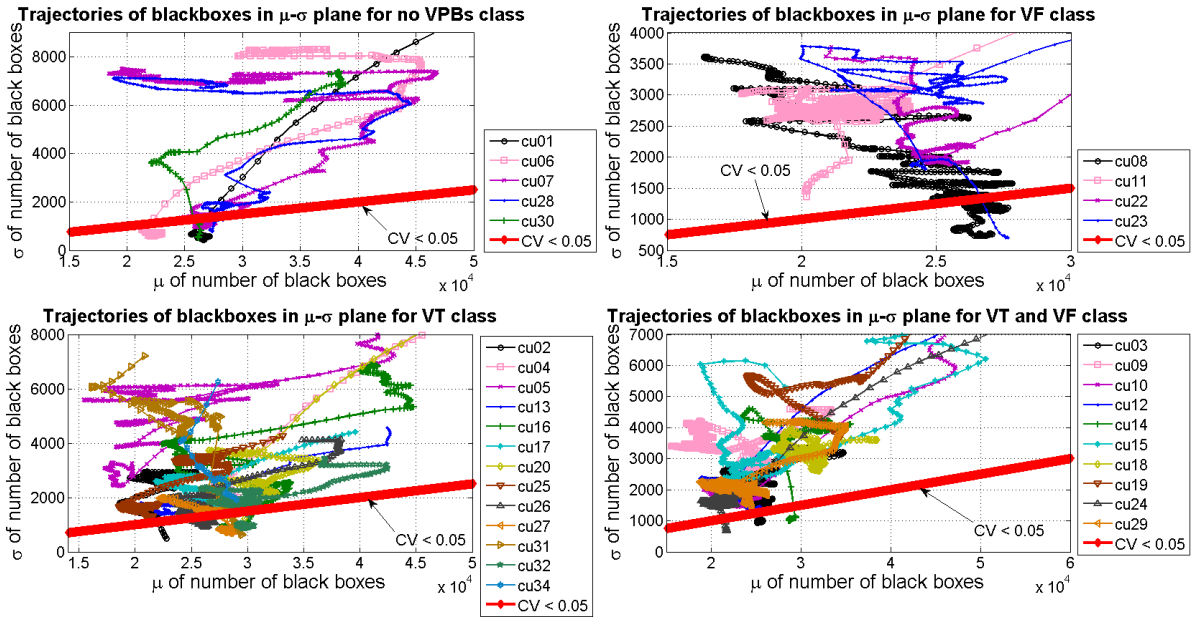

Fig. 12. Time evolution of *CV* trends of box-counting in $\mu$-$\sigma$ plane for arrhythmic subjects.

It is to be noted that even though we analysed two different database of Physionet, the qualitative behaviour of the phase portrait box-counting trends are similar for the healthy (in PTBDB) and the recording before the onset of arrhythmia (in CUDB). Especially, the mean ($\mu$), standard deviation ($\sigma$) and skewness ($\gamma$) trends before the arrhythmia in CUDB subjects are quite similar to that of the healthy subjects in PTBDB. Indeed this is a strength of our analysis that the statistical behaviour of the two database exhibit similar nature for the healthy and arrhythmic subjects before the onset of arrhythmia, although the behaviour is not similar for *CV* and kurtosis ($\beta$) trends which indicates towards developing a statistical marker for the early diagnosis of impending arrhythmia. Therefore, in the following section by combing the *CV* and $\beta$ trends, we formulate a hybrid statistical index with the aim of getting some predictive property to detect almost invisible inter-beat desynchronisation for the VA subjects compared to that of the healthy subjects. Also, previous literatures like [6–8] mostly addressed the detection problem of arrhythmia and as such there is no continuous and online monitoring scenario which is addressed in the present paper. Here, we analysed the whole CUDB database, except three entries *viz.* cu21, cu33 and cu35. These three signals in CUDB were mostly masked with artefact and it was very difficult to extract any useful information from the corrupted the ECG traces. These three entries (cu21, cu33 and cu35) are heavily masked by large and prolonged instances of artefacts before the onset of arrhythmia and therefore no useful information could be extracted for the early diagnosis.

## 5. Formulation of a novel prediction index for early diagnosis of arrhythmia

### 5.1. Deriving the proposed prediction index from PTBDB and analysing arrhythmic subjects in CUDB

From the discussions presented in the foregoing section, it is evident that out of the five statistical indices for characterising the temporal evolution of the underlying desynchronisation phenomenon in ECG, the trends $\beta$ and $CV$ have the most potential predictive power whereas the trends in $\mu$, $\sigma$ and $\gamma$ have more *detection* properties rather than *predictive* properties. It is to be noted that although $CV$ trend has little predictive power for the cases without VPBs its predictive power in the other cases is significant and therefore may not be ignored completely in formulating a predictive index for VA. Based on this logic, we have chosen $\beta$ and $CV$ trends as the main two parameters for formulating a regularized hybrid prediction index $J$.

First we define two thresholds $CV_{th} = 0.05$ and $\beta_{th} = 6$ which denote the upper bounds of the trends in $CV$ and $\beta$ respectively for healthy subjects as shown in Fig. 6 and Fig. 5 respectively. It has also been observed that $CV$ and $\beta$ cross their respective thresholds at sufficient time before the manifestation of arrhythmia for the analysable subjects in CUDB. Therefore by combining them the hybrid prediction index $J$ is formulated as in (6).

$$J = w\frac{CV}{CV_{th}} + (1-w)\frac{\beta}{\beta_{th}}$$ (6)

where, $w$ is a weighting factor bounded within the interval $w \in [0,1]$. The reason for normalising $CV$ and $\beta$ with their respective upper thresholds for healthy subjects is that in absolute terms, their values are markedly different and therefore it is more logical to normalise them to bring them under a uniform platform and thereby eliminating any biased emphasis on one of these two measures. The weighting factor $w$ in essence determines the contributions of $CV$ and $\beta$ towards the final predictive index $J$. As an example $w = 0.5$ allows equal weightage to both the terms, whereas $w = 0$ and $w = 1$ gives full weightage to $\beta$ and $CV$ respectively. The main idea is to find an optimum $w$ that may allow one to compute a $J$, capable of predicting all the VA cases considered here. It is to be noted that since the individual indices are normalised, and the weight of the regularized index is always bounded in $w \in [0,1]$, for healthy subjects, included in this study irrespective of the value of $w$, $J < J_{th} = 1$.

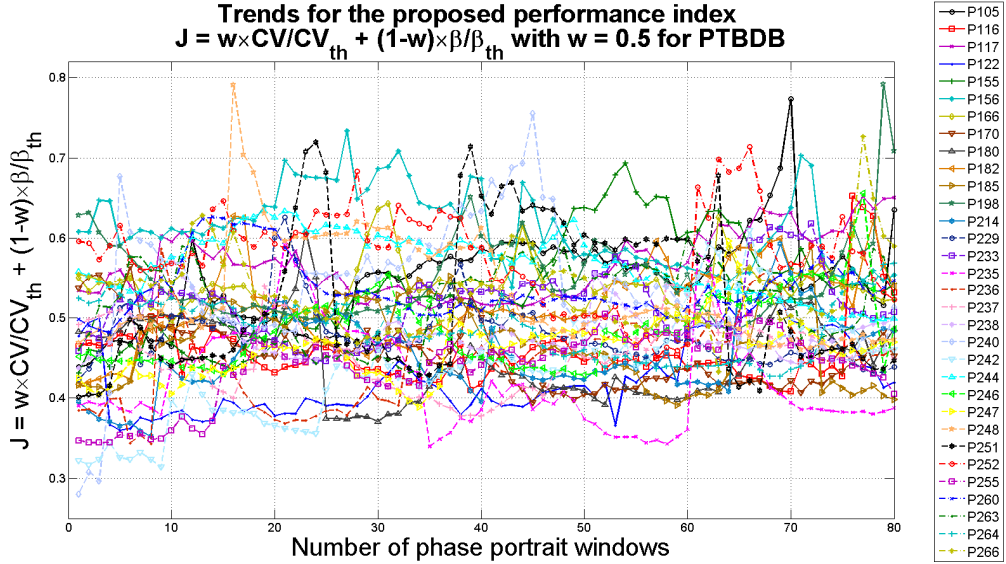

Fig. 13. The hybrid arrhythmia prediction index ($J$) trends of box-counting for healthy subjects with equal impact for $CV$ and $\beta$ trends i.e. $w = 0.5$.

Fig. 13 shows the variation of $J$ for all the healthy subjects with $w = 0.5$ as an example, although the optimum value of $w$ is to be determined yet for the VA subjects. However, if any of $CV$ or $\beta$ crosses their respective threshold signifying the possibility of impending arrhythmia, $J > J_{th} = 1$ condition will result. Therefore it is logical to consider the aforementioned condition as a sufficient condition for predicting arrhythmia. However in reality this is not the case. This can be explained considering an example. Let us consider that $w = 1$, gives full emphasis on the $CV$ trend. Therefore if $CV > CV_{th}$ then $J > J_{th}$ implies a possible impending arrhythmia. In reality it has been observed that $CV$ may cross the threshold $CV_{th}$ in the vicinity of VPBs, even in non-arrhythmic cases. As a result considering $J > J_{th}$ with $w = 1$, certainly leads to a mis-prediction of arrhythmia. This leads to the conclusion that while $J > J_{th}$ is a necessary condition for predicting arrhythmia; it is not sufficient and in principle depends on the value of $w$ chosen. Therefore the main task here is in obtaining an optimal $w$ ($w_{opt}$) – i.e., the optimal balance between the contributions of the trends of $CV$ and $\beta$ – under which $J > J_{th}$ is sufficient for predicting impending arrhythmia with minimum chance of mis-prediction and maximum time for prediction or early diagnosis.

As the first step for deriving the $w_{opt}$, we proceed with patient wise analysis of the trends of $J$ for each class of arrhythmia which are plotted for individual arrhythmic patients (patient number in CUDB mentioned in the respective titles of the subplots) by varying $w$ from 0 to 1 in step of 0.1 as shown in Fig. 14(a)–(d). From the figures it is evident that the trends of $J$ cross the critical threshold of $J_{th} = 1$ at different time instants for the same subject depending on the value of $w$. One approach for choosing $w_{opt}$ may be to select the $w$ for which $J$ crosses $J_{th}$ earliest. Therefore we first record the time of crossing of $J_{th} = 1$ for each choice of $w$ and for each arrhythmic patient from Fig. 14. This is done by first taking the difference between the indices of the phase portrait windows at which $J$ crosses $J_{th}$ and at which the onset of arrhythmia is manifested and subsequently converting that difference in the phase portrait window scale to sec using (5). This time is termed as the prediction time ($T_p$) or the time for early diagnosis. Subsequently, $T_p$ is plotted in Fig. 15 for individual patients while choosing different values of $w$.

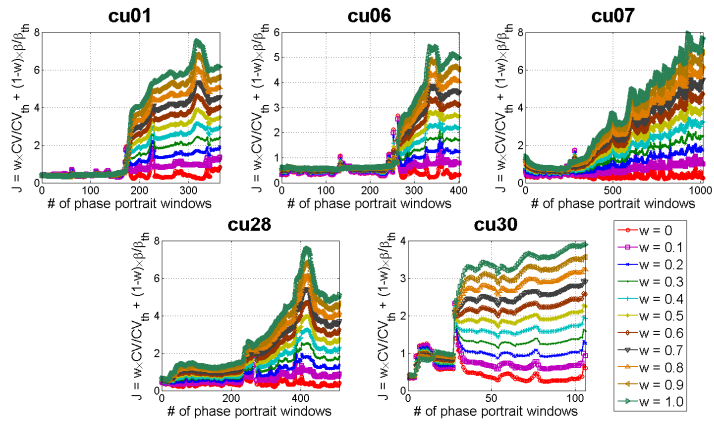

(a)

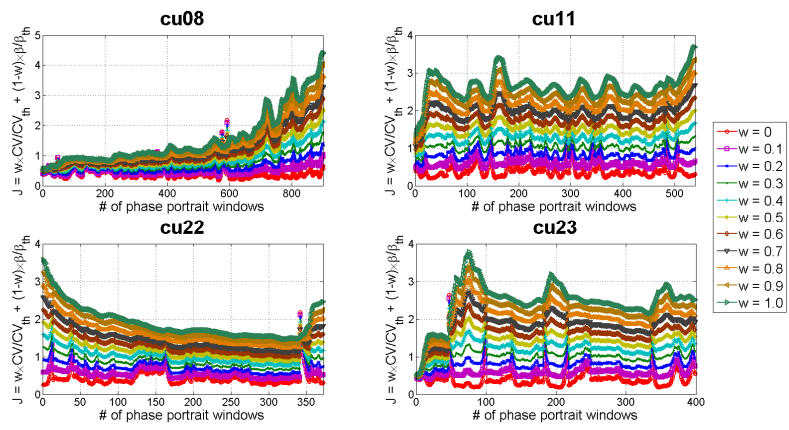

(b)

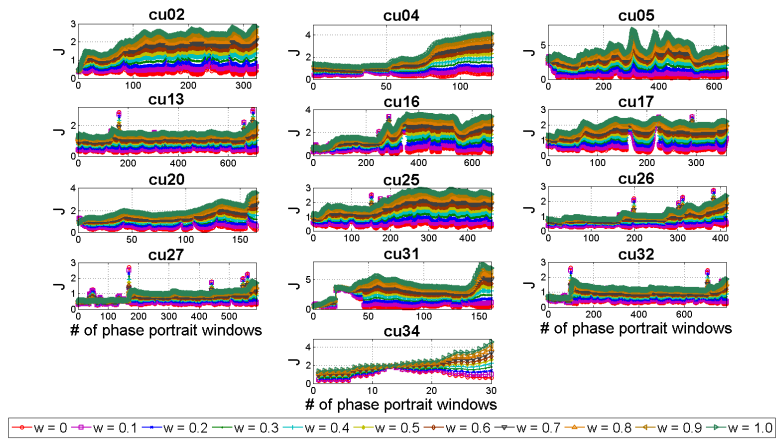

(c)

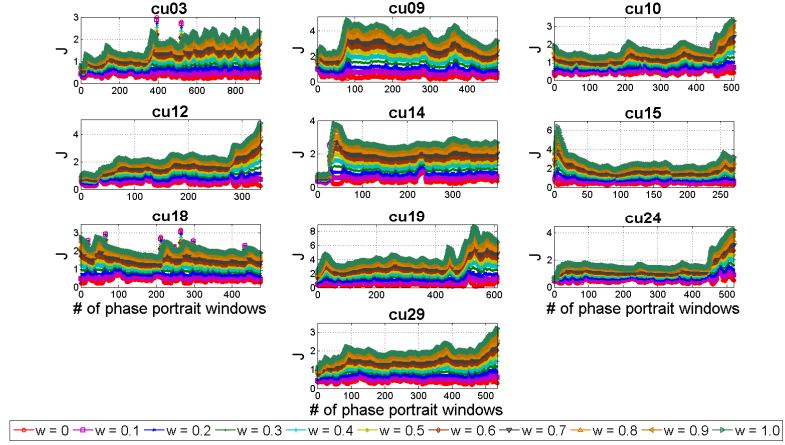

(d)

Fig. 14. Variation in arrhythmia prediction index ($J$) with different weights ($w$) for (a) No-VPB class, (b) VF class, (c) VT class, (d) VT followed by VF class.

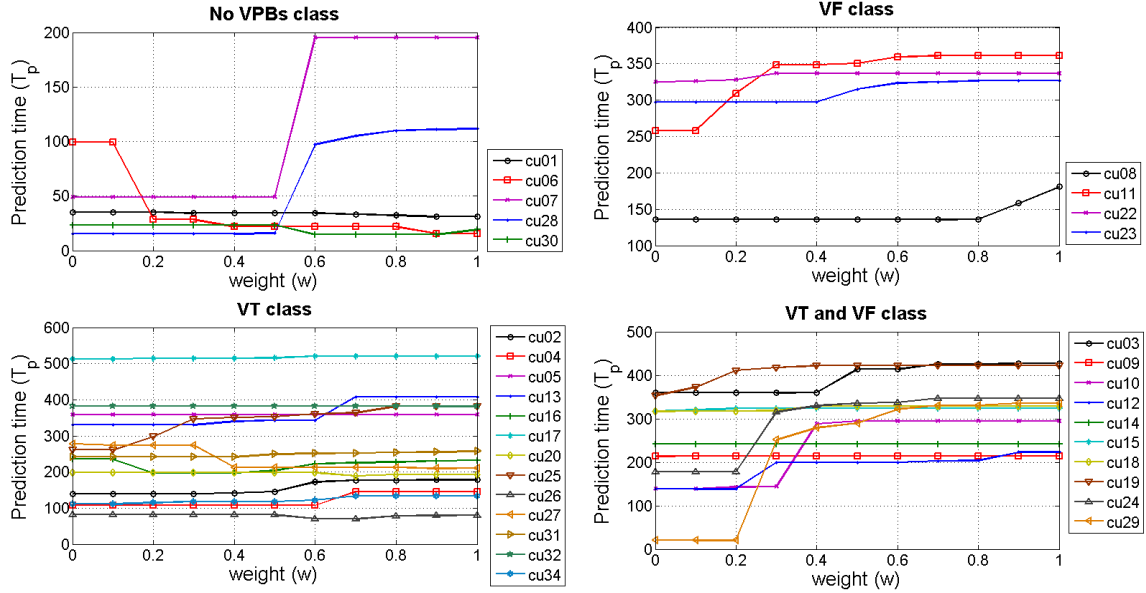

Fig. 15. Variations in prediction time (sec) with different choice of $w$ for the four subclasses of VA.

A close look at Fig. 15 reveals that $T_p$ reaches its maxima at $w=1$ which means only considering the $CV$, as it was lying above $CV_{th}$ from the very beginning in some cases of Fig. 11 and Fig.14. However, it has already been shown in Figs. 11-12 that $CV$ alone is not a reliable predictor for no-VPBs class of arrhythmia. As a result considering $w=1$ although may give maximum prediction time; it is also prone to mis-prediction for no VPB class. Therefore, we introduced another criterion as the total time of misdetection ($T_{mis-det}$) given by the sum of the number of phase portrait windows after the onset of arrhythmia before crossing $J_{th}=1$, while considering all the four groups of arrhythmic subjects. The $T_{mis-det}$ is basically the sum of all negative prediction times, indicating detection of the onset of arrhythmia after its actual occurrence that needs to be minimized for reliable prediction. Accordingly, the new criterion for choosing $w_{opt}$ is to minimize $T_{mis-det}$. This $T_{mis-det}$ is intrinsically different from the average prediction time ($T_{avg-predict}$) for all patients ($N$), corresponding to each selection of $w$, since $T_{mis-det}$ is computed with only those patients ($M \subset N$) showing a negative prediction time.

$$T_{mis-det} = \sum_{i=1}^{M} m_i, \quad T_{avg-predict} = \frac{1}{N}\sum_{i=1}^{N} T_{p_i} \tag{7}$$

The $T_{mis-det}$ is a more informative criterion (for minimization) than $T_{avg-predict}$ (for maximization) as the patients with larger positive prediction time overwhelms small occurrences of misdetection for particular choice of weight $w$. In other words, although the objective of finding $w_{opt}$ could have been posed as maximizing the $T_{avg-predict}$, we have taken an approach of minimizing the total time of misdetection, in terms of number of phase portrait windows that we get before the onset of arrhythmia once $J$ crosses the threshold of $J_{th} = 1$. In Fig. 16 the variation in $T_{mis-det}$ and $T_{avg-predict}$ have been shown with change in $w$ which shows that $T_{mis-det}$ attains its minima at $w = 0.6$. Using this value as $w_{opt}$ in (6) and checking the trends of resulting $J$ in Fig. 14(a)-(d), one can see that long before the actual arrhythmic event $J$ shows a sharp rise followed by crossing of $J_{th} = 1$ and it stays significantly above $J_{th} = 1$ until the arrhythmic event occurs. Physically it signifies that the spread of the trajectories in phase-space normalized by their mean (for *CV*) and also sudden concentration of box counting around its mean (for *β*) both become more dominant over time which indicates towards the build-up of a desynchronisation process, finally leading to arrhythmia.

Optimum weight ($w_{opt}$) selection corresponding to the minimum total misdetection time ($T_{mis-det}$)

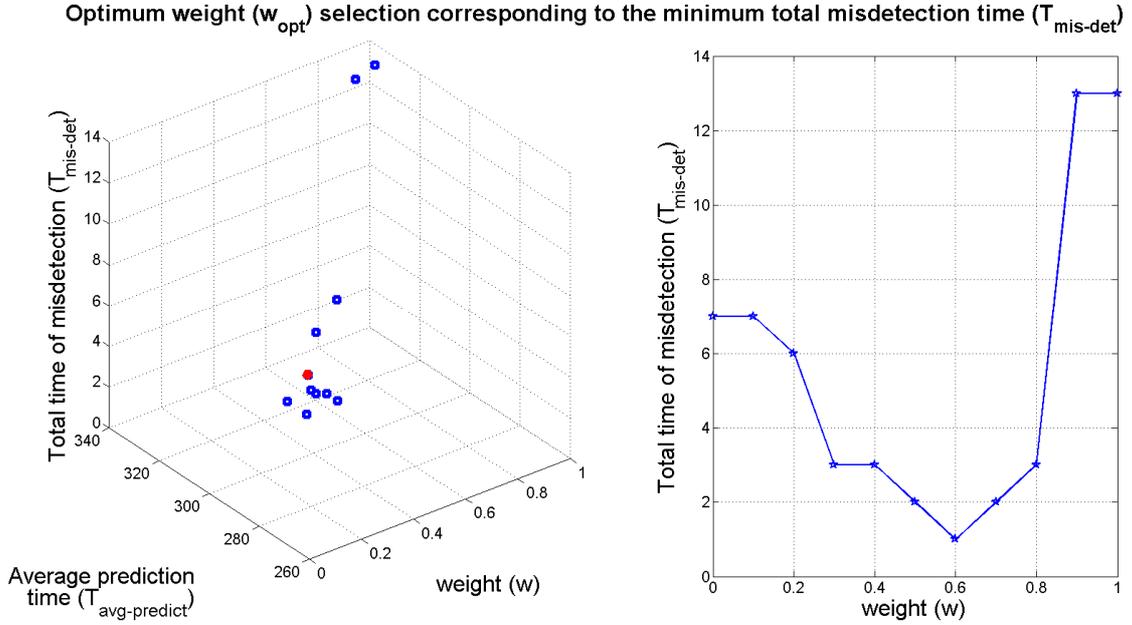

Fig. 16. Obtaining the minima of total time of misdetection ($T_{mis-det}$) in terms of weight ($w$) and average prediction time ($T_{avg-predict}$).

To understand the predictive power of the proposed index $J$, we compute the prediction time $T_p$ for each of the arrhythmic subjects for $w_{opt} = 0.6$ as shown in Table 1. It is evident from Table 1 that for the best prediction, the onset of arrhythmia occurs after 827 ECG beats (cu03) after $J$ crosses $J_{th} = 1$ and for the worst case it is 32 ECG beats (cu01). From the clinical perspective and also as per the study reported in [41] a prediction or early diagnosis of life-threatening arrhythmia on an average of 356 ECG beats (with standard deviation of 192 ECG beats) before its onset allows sufficient time to switch on an alarm and to act for preventive measures i.e. to have a defibrillator ready. Table 1 shows that the equivalent early diagnosis or prediction time (in sec) before the arrhythmia onset can be obtained from the average heart rate of individual patients (calculated from the clinical annotations), with the best and worst case prediction occurring for patient number cu17 and cu30 respectively. Also, it can be observed from Table 1 that for the VA with no VPBs class (cu01, cu06,

cu30) the prediction time is relatively small than the other three classes. This is due to the availability of only a small number of healthy looking ECG beats (i.e. insufficient history for statistical analysis) before arrhythmia in CUDB, as also reported in Fig. 14(a) as the number phase portrait windows.

Table 1: Arrhythmia Prediction Results for Patients in CUDB

| Class of arrhythmia | Patient ID in CUDB | Onset of arrhythmia (the first affected phase portrait) | Prediction time $T_p$ (as # of phase portrait windows) | Equivalent number of beats ($T_p$+33) | Average Heart Rate HR (beats/min) | Equivalent prediction time (in sec) using (5) |
|---|---|---|---|---|---|---|
| VA with no VPBs | cu01 | 170 | -1 | 32 | 56 | 34.3 |
| | cu06 | 254 | 1 | 34 | 93 | 21.9 |
| | cu07 | 341 | 341 | 374 | 115 | 195.1 |
| | cu28 | 230 | 163 | 196 | 121 | 97.2 |
| | cu30 | 29 | 1 | 34 | 140 | 14.6 |
| VF | cu08 | 707 | 338 | 371 | 164 | 135.7 |
| | cu11 | 472 | 470 | 503 | 84 | 359.3 |
| | cu22 | 343 | 343 | 376 | 67 | 336.7 |
| | cu23 | 371 | 355 | 388 | 72 | 323.3 |
| VT | cu02 | 312 | 291 | 324 | 113 | 172.0 |
| | cu04 | 96 | 62 | 95 | 53 | 107.5 |
| | cu05 | 517 | 517 | 550 | 92 | 358.7 |
| | cu13 | 654 | 545 | 578 | 101 | 343.4 |
| | cu16 | 533 | 441 | 474 | 128 | 222.2 |
| | cu17 | 314 | 314 | 347 | 40 | 520.5 |
| | cu20 | 148 | 148 | 181 | 55 | 197.5 |
| | cu25 | 444 | 418 | 451 | 75 | 360.8 |
| | cu26 | 352 | 155 | 188 | 161 | 70.1 |
| | cu27 | 561 | 393 | 426 | 120 | 213.0 |
| | cu31 | 145 | 130 | 163 | 39 | 250.8 |
| | cu32 | 703 | 603 | 636 | 100 | 381.6 |
| | cu34 | 25 | 20 | 53 | 26 | 122.3 |
| VT followed by VF | cu03 | 844 | 794 | 827 | 120 | 413.5 |
| | cu09 | 428 | 428 | 461 | 129 | 214.4 |
| | cu10 | 443 | 443 | 476 | 97 | 294.4 |
| | cu12 | 273 | 237 | 270 | 81 | 200.0 |
| | cu14 | 365 | 338 | 371 | 92 | 242.0 |
| | cu15 | 204 | 204 | 237 | 44 | 323.2 |
| | cu18 | 439 | 439 | 472 | 86 | 329.3 |
| | cu19 | 565 | 565 | 598 | 85 | 422.1 |
| | cu24 | 440 | 417 | 450 | 80 | 337.5 |
| | cu29 | 443 | 423 | 456 | 85 | 321.9 |

Another interesting point to note is that immediately after the occurrence of a VPB, the value of index $J$ changes significantly. This is in conformation with the fact that the VPB firing may be a compensatory mechanism for mitigating the gradual building-up of the desynchronisation process. When the desired synchronisation level is not reached even after firing the first VPB, a series of VPBs are fired in an attempt to attain the desired synchronisation before the arrhythmic event takes place. It is to be noted that in the present study, amongst all the four subclasses of VA considered, three of them exhibit one or more VPBs before the arrhythmia onset. On the other hand the first category of arrhythmia has no VPBs before its onset. Therefore, from the presented simulation study it may be argued that in a continuous monitoring scenario, the crossing of the proposed arrhythmia prediction index threshold of $J_{th} = 1$ with $w_{opt} = 0.6$, predicts a possibly impending arrhythmic event allowing sufficient time for the clinicians to intervene [41] and stop any escalation of it.

### 5.2. Early diagnosis results by the *CV*, kurtosis and the proposed prediction index

Since the both the components of the hybrid prediction index i.e. *CV* and kurtosis was below the respectively threshold for healthy subjects in Fig. 5 and Fig. 6 respectively, hence there is no scope of optimizing their relative importance (*w*) for the healthy group. Although for the arrhythmic subjects different emphasis on these two statistical measures gives different results (Fig. 14) which motivated us to optimize for the relative importance of these two components to find out the best weight *w* yielding minimum time of misdetection as shown in Fig. 16. Also, in order to test the robustness of statistical analysis, we have performed the calculation under a leave one out cross validation (LOOCV) scenario to derive the upper threshold of the *CV*, *β* and prediction index *J* from the healthy subject to test it on the one held out healthy subject and all the VA subjects. From Fig. 5 and Fig. 6 it is evident that the *CV* and kurtosis trends are both below their respective upper thresholds and therefore their weighted sum or the index is always *J*<1 for all the healthy subjects as shown in Fig. 13. In statistical sense the false positive detection is zero in our analysis of 32 healthy subjects. Here, we report the early diagnosis results under the LOOCV approach for the two groups (Positive P to represent VA and Negative N for the healthy subjects) to calculate the true positive (TP), false positive (FP), true negative (TN) and false negative (FN) values. Considering the cross-validation scheme we finally derive the early diagnosis measures like sensitivity (Se) or true positive rate (TPR), specificity (Sp) or true negative rate (TNR), accuracy (Acc), precision or positive predictive value (PPV), negative predictive value (NPV), fall out or false positive rate (FPR), false discovery rate (FDR), miss rate or false negative rate (FNR) and $F_1$ score etc. [42] using the formulae in (8).

$$Se = TP/P = TP/(TP + FN), Sp = TN/N = TN/(FP + TN), \text{Acc} = (TP + TN)/(P + N),$$
$$PPV = TP/(TP + FP), NPV = TN/(TN + FN), FPR = FP/N = FP/(FP + TN), \tag{8}$$
$$\text{FDR} = FP/(FP + TP) = 1 - PPV, FNR = FN/(FN + TP), F_1 \text{ score} = 2TP/(2TP + FP + FN).$$

Table 2: Early diagnosis results using the *CV*, *β* and proposed regularized arrhythmia prediction index *J* under a LOOCV scheme

| Diagnosis measures (%) | Statistical trends of phase portraits | | |
| --- | --- | --- | --- |
| | **CV (*w* = 1)** | **Kurtosis *β* (*w* = 0)** | ***J* (*w* = 0.6)** |
| Sensitivity or TPR | 93.75 | 90.63 | 96.88 |
| Specificity or TNR | 100.00 | 100.00 | 100.00 |
| Accuracy | 96.88 | 95.31 | 98.44 |
| precision or PPV | 100.00 | 100.00 | 100.00 |
| NPV | 94.12 | 91.43 | 96.97 |
| FPR | 0.00 | 0.00 | 0.00 |
| FDR | 0.00 | 0.00 | 0.00 |
| FNR | 6.25 | 9.38 | 3.13 |
| $F_1$ score | 96.77 | 95.08 | 98.41 |

It is to be noted that the main goal of the present work was not detection or classification but early diagnosis or prediction. However, in order to compare the results with standard methods of medical statistics we have provided the performance measures in Table 2 for the *CV* and kurtosis trends, along with their combination i.e. the regularized statistical index, for both the healthy and arrhythmic subjects. Table 2 shows with the proposed statistical index and also with $CV$ and $\beta$ alone, we always get 100% specificity and PPV. Also, the sensitivity, accuracy, NPV and $F_1$ score is best for index $J$, followed by $CV$ and kurtosis which again shows the advantages of the proposed regularized index over its two individual components.

### 5.3. Discussion

Our exploration shows that the sliding window $CV$ and $\beta$ trends of the ECG phase portraits have high predictive value for impending arrhythmia. Subsequently, we hybridize $CV$ and $\beta$ trends for formulating the final regularized predictive index and show that while it consistently stays below a certain threshold for healthy subjects, it crosses that threshold for patients with arrhythmic tendency long before the actual manifestation of VA, and therefore has got some early diagnostic or predictive property for different types of arrhythmic events. The proposed statistical prediction index can be seen as a first step towards having a potential tool for next-generation tele-monitoring of cardiovascular diseases (CVD) like VT and VF subcategories of fatal arrhythmia.

Although the hybrid index formulated in this paper shows good short-term predictive value of VA, it is to be noted that in this analysis we are restricted by the length of ECG data available in the CUDB before the onset of arrhythmia. In several cases the ECG data before the onset of VA is very limited rendering the evaluation of actual performance of the proposed index difficult. Also, the available data in CUDB did not provide information regarding the long-term clinical status of the patients (e.g., the risk-level, medication etc.) which may have some implications on the patient-centric arrhythmic property and hence the performance of the proposed index. Also, for a specific choice of ECG lead, the morphology of the beats remains almost consistent during sinus rhythm or at rest for healthy subjects but even in such cases the heart rate and the nature of ECG phase portraits may change on mood, anxiety level or physical activity. However, in general it is observed that longer the ECG data before VA onset better is the predictive property of the proposed index, since it derives the threshold-crossing criteria through statistical analysis. We believe that prospective analysis with large cohort is needed with long-term ECG recording to evaluate the true potential of the proposed index. The results presented here is the first step towards that direction and although analysed with only short ECG data its performance seems promising.

### 6. Conclusion

In this paper, we propose a novel statistical index ($J$) for the early diagnosis or prediction of ventricular arrhythmia, in particular four sub-classes *viz.* no VPBs, VT, VF and VT followed by VF, using the phase-space reconstruction method of long-term ECG time-series. We found that while $CV < 0.05$, $\beta < 6$ and consequently $J < 1$ signify healthy condition. $J > 1$ with $w_{opt} = 0.6$ predict an impending arrhythmia and therefore has a potential to be applied as an effective tool for predicting fatal arrhythmia. The worst and the best case early diagnosability or predictability of 32 and 827 ECG beats respectively, with an average prediction time of 356 ECG beats (having standard deviation of 192 beats), tested over 32 arrhythmic patients are deemed as sufficient time for taking preventive actions in clinical settings [41]. The proposed hybrid prediction index is verified with a LOOCV strategy and has been shown to give better early diagnosis result over that with $CV$ and $\beta$ alone, with a 96.88% sensitivity, 100% specificity, and 98.44% accuracy. However, a large-scale patient trial is needed to confirm the predictive power of the proposed arrhythmia prediction index. Further study can be directed towards quantifying the correlation between the time/frequency of occurrence of the VPBs and first threshold crossing time and the actual onset of arrhythmia.

### Acknowledgement

This work was supported by the E.U. ARTEMIS Joint Undertaking under the Cyclic and person-centric Health management: Integrated appRoach for hOme, mobile and clinical eNvironments – (CHIRON) Project, Grant Agreement # 2009-1-100228.